%% file: main.tex
\documentclass[10pt,twocolumn,letterpaper]{article}
\pdfoutput=1

\usepackage{cvpr}
\usepackage{times}
\usepackage{epsfig}
\usepackage{graphicx}
\usepackage{amsmath}
\usepackage{amssymb}

\usepackage[pagebackref=true,breaklinks=true,letterpaper=true,colorlinks,bookmarks=false]{hyperref}

\cvprfinalcopy 

\usepackage{aux/custom}

\makeatletter
\newcommand{\printfnsymbol}[1]{%
  \textsuperscript{\@fnsymbol{#1}}%
}
\makeatother

\begin{document}

\title{torchgpipe: On-the-fly Pipeline Parallelism for Training Giant Models}



\author{Chiheon Kim\textsuperscript{1}\thanks{Contributed equally.} \qquad Heungsub Lee\textsuperscript{1}\footnotemark[1]
\qquad Myungryong Jeong\textsuperscript{1} \qquad Woonhyuk Baek\textsuperscript{1} \\
Boogeon Yoon\textsuperscript{1} \qquad Ildoo Kim\textsuperscript{1} \qquad Sungbin Lim\textsuperscript{2}\thanks{This work was done while Sungbin Lim was at Kakao Brain.} \qquad Sungwoong Kim\textsuperscript{1}\\
\textsuperscript{1}Kakao Brain \qquad \textsuperscript{2}UNIST \\
\textsuperscript{1}{\tt\small \{chiheon.kim,heungsub.lee,myungryong.jeong,wbaek,eric.yoon,ildoo.kim,swkim\}@kakaobrain.com} \and \textsuperscript{2}{\tt\small sungbin@unist.ac.kr}
}

\maketitle

\begin{abstract}
We design and implement a ready-to-use library in PyTorch for performing micro-batch pipeline parallelism with checkpointing proposed by GPipe \cite{huang2019gpipe}. In particular, we develop a set of design components to enable pipeline-parallel gradient computation in PyTorch's define-by-run and eager execution environment. We show that each component is necessary to fully benefit from pipeline parallelism in such environment, and demonstrate the efficiency of the library by applying it to various network architectures including AmoebaNet-D \cite{real2019regularized} and U-Net \cite{RonnebergerFB15}. Our library is available at \url{https://github.com/kakaobrain/torchgpipe}.
\end{abstract}


\input{sections/1_introduction.tex}
\input{sections/2_pipeline-parallel.tex}

\input{sections/3_torchgpipe.tex}
\input{sections/4_experiments.tex}
\input{sections/5_conclusion.tex}

{\small
\bibliographystyle{ieee_fullname}
\bibliography{aux/references}
}

\end{document}

%% file: sections/1_introduction.tex
\section{Introduction}
\label{sec:introduction}

In recent years, deep learning has seen significant growth, driven by several methodologies which enable the training of deep neural networks (DNNs) in a scalable way and by development of more powerful hardwares. It is observed that increased capacity of DNN effectively has improved the performance. For example, AmoebaNet-B \cite{real2019regularized} scaled with GPipe \cite{huang2019gpipe} has 557 million parameters and has achieved top-1 accuracy 84.4\% which was state-of-the-arts result at the time, and GPT-2 \cite{radford2019language} is a Transformer-based \cite{vaswani2017attention} language model which has 1.5 billion parameters (see Figure 1 of \cite{huang2019gpipe} for the effect of model scaling). However, training such a massive model is very resource intensive. One can mitigate this issue by reducing the size of the model without losing the performance by pruning the model \cite{han2015learning, alvarez2016learning}, designing more efficient architectures \cite{howard2017mobilenets, tan2019efficientnet}, architecture search under resource constraints \cite{cai2018proxylessnas}, and many more. 

We may wonder a rather direct approach is possible: \emph{can we train a massive model fast enough, given a large pool of devices?} One obstacle is that common optimization techniques to train a neural network are sequential in nature. Those algorithms repeatedly compute the gradient of the loss with respect to the given mini-batch at a time and update the model parameters using the gradient. With abundant computational resource, data parallelism \cite{krizhevsky2012imagenet} is commonly used to speed up the overall optimization procedure by dividing the mini-batch into micro-batches and delegating per micro-batch computation to available devices. With careful hyperparameter tuning, this effectively reduce the training time up to a certain size of mini-batch which may depend on model, optimization algorithm, and data \cite{goyal2017accurate, shallue2018measuring}. One drawback of data-parallel training is that devices hold their own version of network for executing the subdivided task, and network parameters must be synchronized after each parameter update. This may induce heavy communication load when there are lots of parameters to synchronize.

Note that data parallelism is not applicable when the model is so big that it is impossible to compute gradient even when a single data point is fed into the network. Model parallelism \cite{dean2012large} is a method for training such a massive model, which partitions the model into several pieces and places them on different devices. Each device only computes a small part of the model, and updates only the parameters in that part. However, model parallelism suffers from its underutilization behavior. Since most neural networks consist of sequence of layers, the device holding the later part of the model must wait until computation in devices holding earlier parts of the model. 

Another possible solution is to use gradient checkpointing \cite{chen2016training} which saves memory by only storing the subset of activation maps and re-computing the discarded activation maps when necessary. Obviously, this requires certain part of the model be computed twice and overall training time would be increased. 

It is benefitting to combine different types of parallelization strategies \cite{krizhevsky2014one, pmlr-v80-jia18a, shazeer2018mesh, huo2018decoupled, harlap2018pipedream, huang2019gpipe, guan2019xpipe}, and recent lines of research questions how to find an optimal strategy \cite{jia2018beyond, mirhoseini2017device, mirhoseini2018a, zhou2019gdp}. Among them, pipeline parallelism a way to accelerate neural network training by combining model parallelism with data pipelining, either in synchronous way as in GPipe \cite{huang2019gpipe} or in asynchronous way as in  \cite{huo2018decoupled}, PipeDream \cite{harlap2018pipedream}, and XPipe \cite{guan2019xpipe}. We remark that gradient checkpointing (also called re-materialization) is further combined in GPipe to allow training even bigger models.

In this paper, we design and implement {\torchgpipe}, a ready-to-use library for GPipe in PyTorch \cite{paszke2017automatic}. In particular, we develop a set of design components for optimized pipeline-parallel computations in PyTorch's \emph{define-by-run} and \emph{eager execution} environment. We show that each component is necessary to fully benefit from pipeline parallelism in such environment, and demonstrate the efficiency of {\torchgpipe} by conducting the speed and memory benchmarks on AmoebaNet-D \cite{real2019regularized} and U-Net \cite{RonnebergerFB15} when trained with the library.

The rest of the paper is organized as follows.
In \autoref{sec:pipeline-parallelism}, we discuss how the forward and backward passes can be decomposed into subtasks (under certain assumptions), describe the device placement strategy of micro-batch pipeline parallelism, and demonstrate what the desired order of execution per device is. In \autoref{sec:torchgpipe}, we discuss complications for achieving the optimal timeline of pipeline parallelism in PyTorch and explain how {\torchgpipe} resolves them. Additionally, we relax the assumption that the model is sequentially composed, and provide a way for expressing models with long skip connections so that pipeline parallelism still applies without giving up the efficiency. Then, we demonstrate that the optimization components suggested in the paper are essential for the performance, and evaluate the performance of the proposed library in \autoref{sec:experiments}.

%% file: sections/2_pipeline-parallel.tex
\section{Pipeline Parallelism}
\label{sec:pipeline-parallelism}

Suppose that we have a neural network which is represented as a composition of sequence of subnetworks. Let us denote the subnetworks by $f^1, \cdots, f^n$ with parameters $\theta^1, \cdots, \theta^n$ and let the full network be 
\[
    f = f^n \circ f^{n-1} \circ \cdots \circ f^1,
\]
parameterized by $\theta = (\theta^1, \cdots, \theta^n)$. For clarity, we call $f^j$ \emph{the $j$th partition of $f$} and assume that the parameters of partitions are mutually disjoint. 

When training the network, gradient-based methods such as stochastic gradient descent requires computing the outcome $f(x)$ of the network given a mini-batch $x$ of training data and the corresponding loss, and the gradient $g$ of the loss with respect to the network parameter $\theta$. Those two stages are called forward and backward pass, respectively. 

Since $f$ is sequentially composed, in forward pass $f(x)$ can be computed by letting $x^0 = x$ and sequentially applying the partitions as $x^j = f^j(x^{j-1})$ for $j=1,\cdots,L$. Furthermore, if $x$ consists of $m$ smaller batches $x_1, \cdots, x_m$ called \emph{micro-batches}, computing $f(x)$ dissolves into tasks $F_{i,j}$ where $x^0_i = x_i$ and 
\begin{equation}
\tag{$F_{i,j}$}
x_i^j \ot f^j(x_i^{j-1})
\end{equation}
for $i=1,\cdots,m$ and $j=1, \cdots, n$, assuming that $f$ does not involve any intra-batch computation. One prominent exception for this is batch normalization \cite{pmlr-v37-ioffe15}\footnote{Applying pipeline parallelism to a network with batch normalization is feasible while the computation is not identical anymore. Indeed, this discrepancy also exists in data-parallel training scheme and it may results in degradation of the result.}. The loss is obtained by aggregating $x_i^n = f(x_i)$ and evaluating the loss function on them.

In a similar fashion, backward pass is decomposed into tasks $B_{i,j}$ where $dx_i^{n}$ is the gradient of the loss with respect to $x_i^n$ and
\begin{equation}
\tag{$B_{i,j}$}
\begin{aligned}
dx_i^{j-1} &\ot \partial_x f^j (dx_i^{j}) \\
g_i^{j} &\ot \partial_{\theta^j} f^j (dx_i^j)
\end{aligned}
\end{equation}
for $i=1,\cdots,m$ and $j=1,\cdots,n$. Here
\[
    \partial_x f^j : v \mapsto v^T \cdot \left.\frac{d f^j}{d x}\right|_{x=x_i^{j-1}}
\]
is a function which does backward propagation (also known as vector-Jacobian product) through the partition $f^j$, and $\partial_{\theta^j} f^j$ is defined likewise. As a result, we get the gradient of the loss with respect to $\theta^j$ by summing $g_i^j$ over $i$'s.

\begin{figure*}
    \centering
    \hspace{0.02\textwidth}
    \begin{minipage}[c][340pt][t]{0.40\textwidth}
        \vspace*{\fill}
        \centering
        \includegraphics[width=0.95\textwidth]{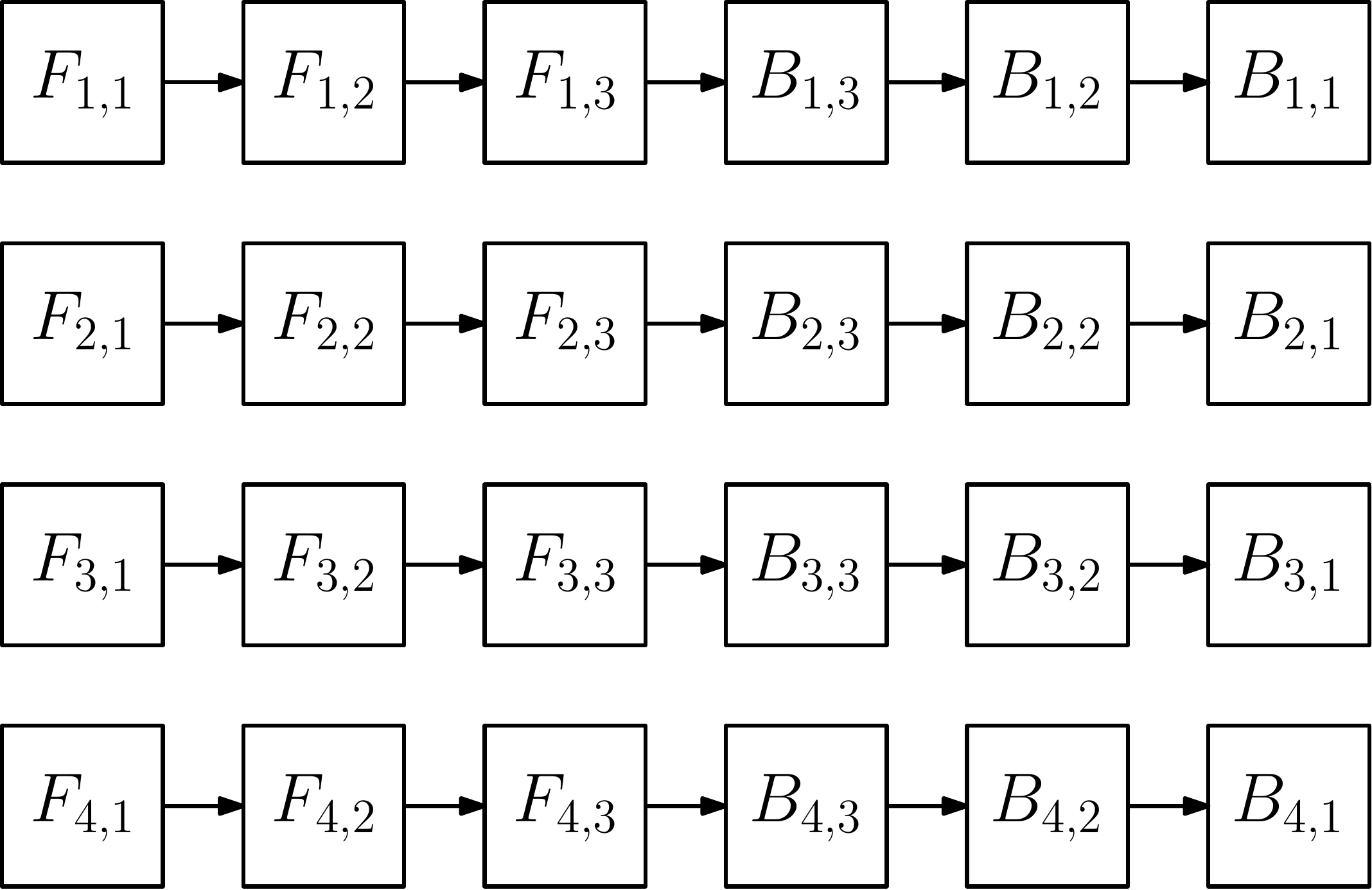}
        \caption{Minimal dependency graph for forward and backward pass.}
        \label{fig:dependency}\par\vfill
        \vspace{10pt}
        \includegraphics[width=\textwidth]{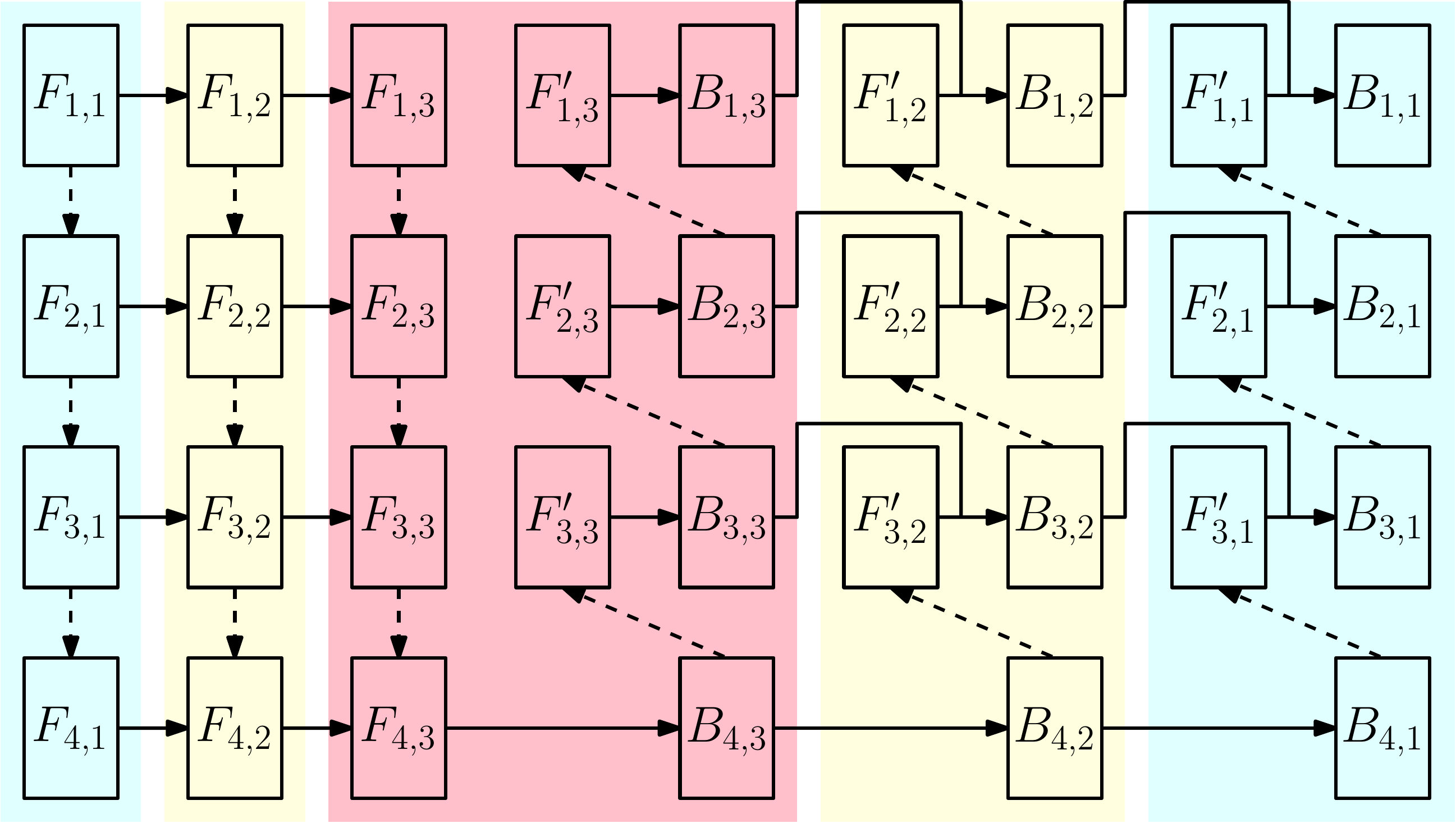}
        \caption{Dependency graph for pipeline parallelism with checkpointing. Colors denote the devices that tasks are computed in.}
        \label{fig:pp-with-checkpointing}
    \end{minipage}
    \hspace{0.03\textwidth}
    \begin{minipage}[c][292pt][t]{0.53\textwidth}
        \vspace*{\fill}
        \centering
        \includegraphics[width=\textwidth]{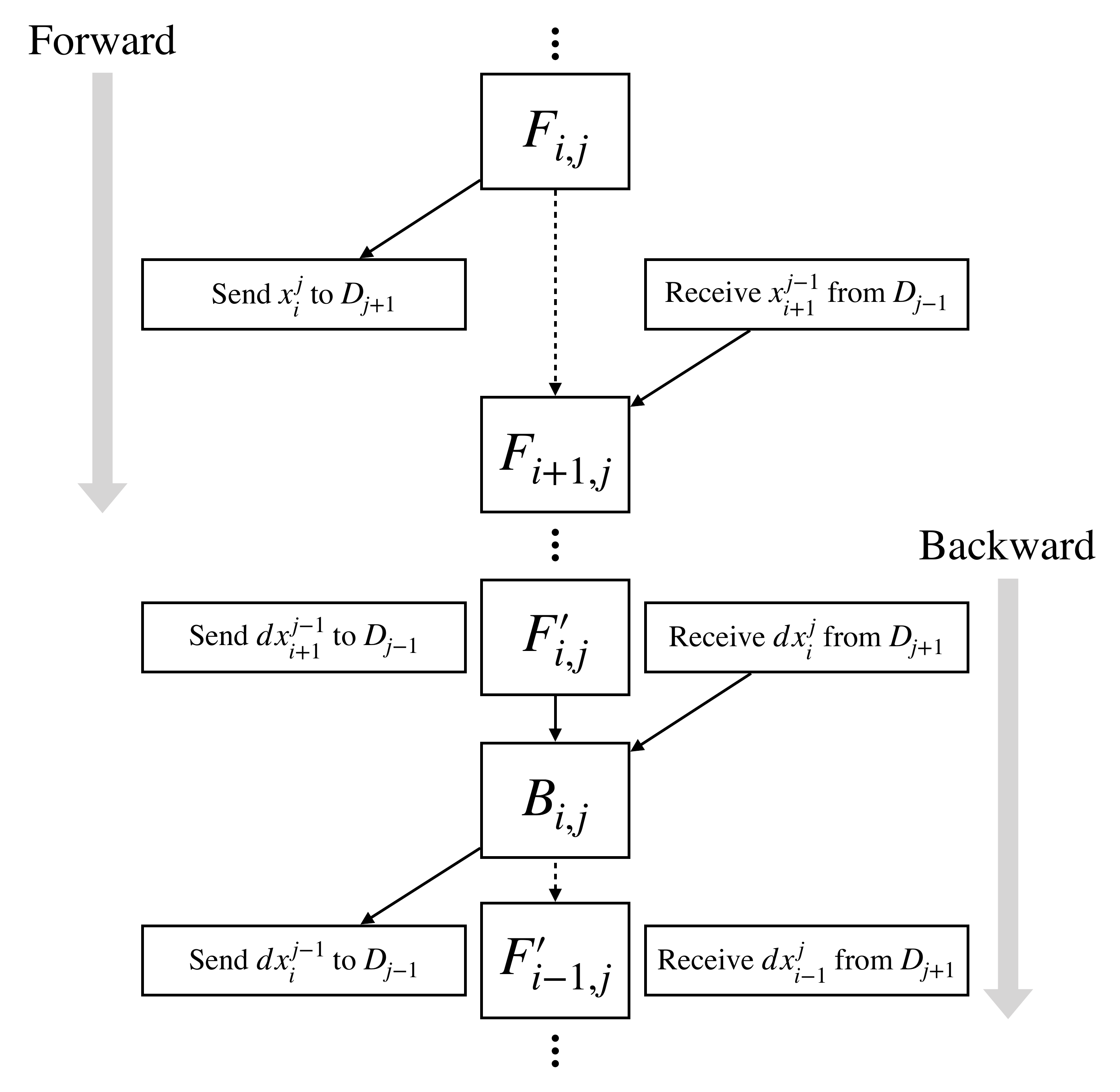}
        \caption{The execution order that $j$th device must follow.}
        \label{fig:kernel-order}
    \end{minipage}
\end{figure*}

Note that there are data dependencies between tasks. For example, $F_{i, j}$ requires $x_i^{j-1}$ which is only available after $F_{i, j-1}$, hence $F_{i, j-1}$ must be completed before starting $F_{i, j}$ and the same applies for $B_{i, j}$ and $B_{i, j+1}$. \autoref{fig:dependency} shows the full dependency graph in the case of $m=4$ and $n=3$.

Given the set of tasks $\{F_{i, j}\}$ and $\{B_{i, j}\}$ and a pool of devices which can work in parallel, different parallelization strategies have their own rule to assign tasks to devices. Each device computes one or more assigned tasks as soon as the dependencies are resolved. In the setting above, all dependencies are among the tasks with the same micro-batch index $i$. Hence, one can effectively parallelize the tasks by assigning tasks with different micro-batch indices to different devices --- which is data parallelism.

\subsection{Dependency Graph of GPipe}
\label{subsec:gpipe-dependency}

Pipeline parallelism's strategy is to assign tasks with respect to the partition index $j$ so that $j$th partition entirely lies in the $j$th device. In addition to this, it is enforced that $F_{i, j}$ must be completed before executing $F_{i+1, j}$ and $B_{i, j}$ must be completed before executing $B_{i-1, j}$. 

In addition to the micro-batch pipelining, GPipe \cite{huang2019gpipe} further reduces the memory requirement by utilizing gradient checkpointing for each $B_{i, j}$. Since $j$th device executes $B_{i,j}$ one at a time, only the activation maps obtained from $F_{i,j}$ are needed to complete $B_{i,j}$. By recomputing the forward pass $F_{i,j}$ right before executing $B_{i,j}$, memory consumption is reduced by a factor of $m$. Moreover, the re-computation can take place while the device is waiting for $B_{i,j+1}$ being done. This is summarized in \autoref{fig:pp-with-checkpointing}, where dashed arrows denotes the execution order between independent tasks induced by the micro-batch order, and $F_{i, j}'$ denotes the re-computation of $F_{i,j}$.

We remark that re-computations for the last micro-batch, \emph{i.e.}, $F_{m, j}'$ for $j=1,\cdots,n$ are unnecessary. This is because that on $j$th device the last task in the forward pass is $F_{m, j}$, so discarding intermediate activations of it in forward pass and re-computing them in the beginning of backward pass has no effect of reducing memory, only slowing down the pipeline. For this reason, $F_{m, j}'$ is omitted from the graph.

\subsection{Device-wise Execution Order}
\label{subsec:timeline}

To summarize, in pipeline parallelism (with checkpointing) each device is assigned with a set of tasks with the prescribed order. Each device will execute the given tasks one-by-one as soon as cross-device dependencies are met. However, there is a missing component in this picture --- data tranfer between the devices. For illustration, the full execution order that device $j$ must follow is shown in \autoref{fig:kernel-order}. Here data transfer operations are explicitly denoted as `receive' and `send' for emphasis.


%% file: sections/3_torchgpipe.tex
\section{{\torchgpipe}: A PyTorch Library for GPipe}
\label{sec:torchgpipe}

{\torchgpipe} is a PyTorch library for micro-batch pipeline parallelism with checkpointing, as known as GPipe. The library provides a simple way to apply GPipe to a generic sequential module written in PyTorch. The usage of {\torchgpipe} resembles that of the data parallel module of PyTorch --- just wrap your model with the wrapper.

Users must specify the number of micro-batches $m$ and how consecutive layers form $n$ partitions. Here we remark that even though we simplified our assumption to that the model is a sequence of partitions, it is strictly required in {\torchgpipe} that the model is a sequence of \emph{layers} to give flexibility for users how to split the model. {\torchgpipe} will assume that each layer is a non-divisible, black-box, and referentially transparent\footnote{This is required especially for checkpointing: referential transparency ensures that recomputation is identical to the computation done in the forward pass.} algorithm. 

For convenience, the library provides the submodule \cmtt{torchgpipe.balance} which computes a partition whose pairwise resource discrepancy is small, where resource consumption is computed by profiling. Specifically, we used the algorithm from \cite{barany2015block}.

As {\torchgpipe} is built on PyTorch equipped with CUDA backend, we will often assume that devices are NVIDIA GPU throughout this section. Nevertheless, the underlying principle of the library applies in general for implementing pipeline parallelism any eager execution environments.

\subsection{Complications in PyTorch}
\label{subsec:obstacles}
Our primary concern is efficiency. As we discussed in \autoref{subsec:timeline}, in order for pipeline parallelism to work as desired, the tasks must be assigned to each device in the correct order. There are several complications to achieve this in PyTorch. 

First of all, kernels are issued to each device on-the-fly due to PyTorch's define-by-run style and its eager execution behavior (as opposed to in construct-and-run type frameworks). Hence, one must design the host code carefully so not only that device-bound tasks are issued in the correct order within each device, but also that execution of the tasks on devices (asynchronous to CPU) are not delayed due to the Python interpreter failing to request it ahead of the time. This kind of delay may happen when some of the tasks are CPU-intensive or involve a lot of cheap kernel calls. As a solution, {\torchgpipe} introduces deterministic clock-cycle which gives the total ordering of the tasks.

Secondly, the computation graph for backward pass is constructed dynamically during the forward pass in PyTorch. In other words, \emph{``it avoids ever materializing a ``forward graph'', recording only what is necessary to differentiate the computation.''} \cite{paszke2017automatic} Since PyTorch does not record the forward computation graph nor maintain a gradient tape, the automatic differentiation (autograd) engine of PyTorch does back-propagation solely with respect to the graph. It implies that autograd engine may not run exactly in the reverse order of execution as in the forward pass, unless enforced by the structure of the graph. To deal with this, we develop a pair of primitive functions called `fork' and `join' to create explicit dependencies on the fly in the backward computation graph.

Thirdly, communication between several devices can cause two-way synchronization, if not carefully managed. This may cause under-utilization since sender may wait to synchronize with the receiver even when there is no explicit dependency between the copy and next task in queue, or vice versa. {\torchgpipe} avoids this issue by using non-default CUDA streams so that copies would never block computations unless the computation must wait for the data.

Lastly, {\torchgpipe} attempts to relax the restriction of micro-batch pipeline parallelism that model must be sequential. Although any neural network can be written in a sequential form in principle, this requires knowing the entire computation graph ahead of the time which is not the case in PyTorch. In particular, if there is a tensor which skips from a layer in device $j'$ to another layer in device $j > j'+1$, the tensor will be copied to all devices in between since {\torchgpipe} cannot know it ahead.
To circumvent this issue, we design an interface to signify which intermediate tensors are skipped and which layers use them.

\subsection{Optimization Components}
\label{subsec:implementation}
In the remainder of this section, it is explained how the components of {\torchgpipe} are designed and why each of them is essential for performance.

\subsubsection{Forward Dependency: Deterministic Clock-cycle}
\label{subsubsec:deterministic-clock-cycle}
As we discussed in \autoref{subsec:obstacles}, the total ordering of tasks is determined by the host code in the forward pass. Each device implicitly understands the dependency between tasks by the order they are assigned by CPU. Ideally, if tasks could be assigned to devices with no cost, CPU may assign tasks to devices in any order as long as the ordering within device is correct. However, this assumption is not realistic enough, as launching kernels on a GPU is not free for CPU, memory transfer between GPUs may require synchronization, or a task is CPU-intensive. For this reason, we minimize the delay coming from CPU by sorting all tasks by the distance to $F_{1,1}$.

\begin{algorithm}
\For{$k$ from $1$ to $m+n-1$}{
    \For{$i, j$ such that $i + j - 1 = k$}{
        \If{$j > 1$}{
            Copy $x_i^{j-1}$ to device $j$.
        }
    }
    \For{$i, j$ such that $i + j - 1 = k$}{
        Execute $F_{i,j}$.
    }
}
\caption{Deterministic clock-cycle}
\label{alg:clock-cycle}
\end{algorithm}

We call this \emph{deterministic clock-cycle} (\autoref{alg:clock-cycle}). In the algorithm, CPU executes the clock cycles starting from the counter $k=1$ to $k=m+n-1$. In $k$th clock cycle, all copy kernels for data needed to execute tasks $F_{i,j}$ where $i+j-1=k$ are first issued, and then the computation kernels for executing the tasks are registered to corresponding devices (which can be safely multithreaded since tasks in the same clock cycle are independent).

\subsubsection{Backward Dependency: Fork and Join}
\label{subsubsec:fork-join}

\begin{figure}
\center
    \includegraphics[width=0.5\textwidth]{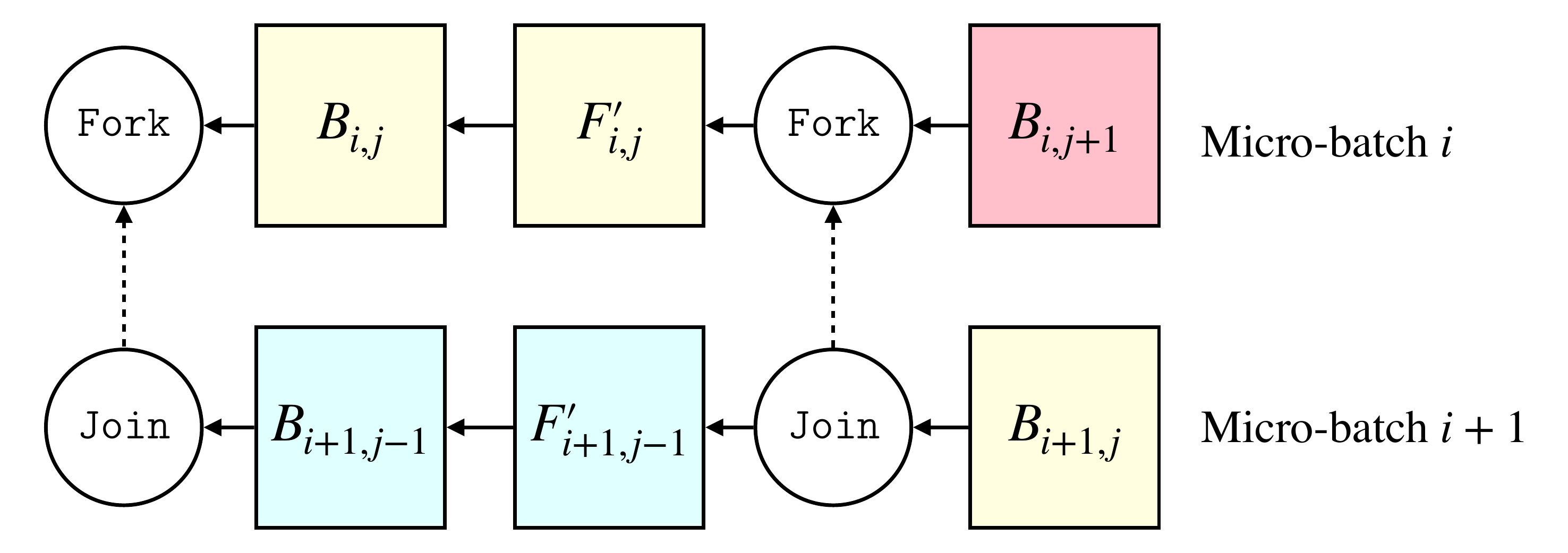}
    \caption{The backward computation graph with \cmtt{Fork} and \cmtt{Join}. Different colors correspond to different devices. Arrows are drawn according to the \emph{direction in backward computation graph} and these relations are constructed during the forward pass. Here the virtual depedency of $F_{i,j}'$ on $B_{i+1, j}$ is created via \cmtt{Fork} and \cmtt{Join}, which is illustrated by dashed arrows.}
    \label{fig:fork-and-join}
\end{figure}

Suppose now that we run a forward pass according to the deterministic clock-cycle. The resulting computation graph for backward will look rather like \ref{fig:dependency} than \ref{fig:pp-with-checkpointing}, even when the forward tasks $F_{1,j},\cdots,F_{m,j}$ on device $j$ were executed in order. From such a graph, autograd engine of PyTorch would never know that $B_{i+1, j}$ must be executed before $B_{i, j}$, and this messes up the timeline of the backward pass. For this reason, virtual dependencies (dashed arrows in \autoref{fig:pp-with-checkpointing}) must be explicitly drawn during the forward pass.

We design a pair of primitive functions called \cmtt{Fork} and \cmtt{Join} to express such dependency. Basically, \cmtt{Fork} is the autograd function mapping a tensor $x$ to the pair $(x, \varnothing)$ where $\varnothing$ is an empty tensor\footnote{In principle, the tensor which indicates the virtual dependency can be arbitrary. We chose to use the empty tensor for this, however, to remove any unnecessary computation caused by the tensor such as gradient accumulation in PyTorch.}, and \cmtt{Join} is the autograd function mapping a pair $(x, \varnothing)$ to the tensor $x$. Now, dependency of $F_{i+1, j}$ upon $F_{i, j}$ (which translates to the dependency of $B_{i, j}$ upon $B_{i+1, j}$ in the backward computation graph) can be expressed as
\[
\begin{aligned}
( x_i^j, \varnothing ) &\ot \text{\cmtt{Fork}}(x_i^j) \\
x_{i+1}^{j-1} &\ot \text{\cmtt{Join}}( x_{i+1}^{j-1}, \varnothing ).
\end{aligned}
\]
See \autoref{fig:fork-and-join} for illustration.

\begin{figure*}[b]
\center
    \includegraphics[width=\textwidth]{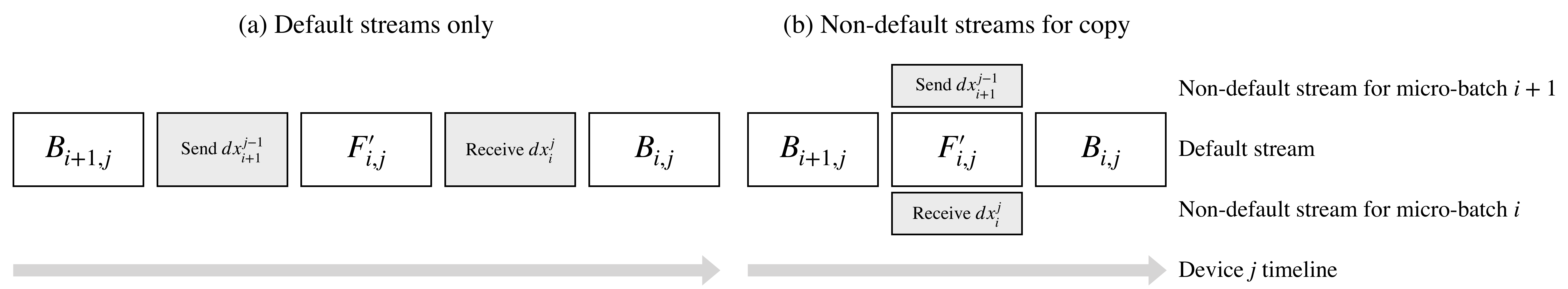}
    \caption{Timeline of device $j$ with or without non-default streams for copy. (a): If only default streams are used, copy kernels may block computation kernels (and vice versa) until the copy is completely finished. (b): With copy streams, computation can happen in concurrent with sending or receiving data from other devices.}
    \label{fig:copy-streams}
\end{figure*}

\subsubsection{Concurrent Copy and Computation: Streams}
\label{subsubsec:copy-streams}

PyTorch issues every device-bound kernels to the default stream, unless it is specified otherwise. Stream is a device-bound sequence of kernels that is executed in order. Kernels in the same stream are guaranteed to be executed in the prescribed order, but kernels in different streams can be interleaved, and even can overlap when possible. In particular, nearly all CUDA devices with compute capability 1.1 and higher support concurrent copy and execution: data transfer between devices can always overlap with kernel execution (see section 4.5.1.5 of \cite{nvidia2007cuda}).

{\torchgpipe} registers every copy kernel to non-default streams while keeping computation kernels on the default stream. This allows the device $j$ processing $F_{i, j}$ in concurrent with sending $x_{i-1}^j$ to the device $j+1$ and/or receiving $x_{i}^{j-1}$ from the device $j-1$. Moreover, each device uses different streams for each micro-batch. Since there is no true dependency between different micro-batches, this use of streams is safe and this allows copies to occur as fast as possible. See \autoref{fig:copy-streams} for illustration.

\subsubsection{Autograd Functions with Shared Memory}
\label{subsubsec:checkpointing}
So far in this section, we did not discuss how to schedule re-computation tasks $F_{i,j}'$ when gradient checkpointing is in use. It must be scheduled in prior to the back-propagation task $B_{i, j}$ upon completion of $B_{i+1, j}$. This must be encoded in the computation graph as well for autograd engine. Indeed, PyTorch supports such functionality via an in-house autograd function for checkpointing.

Checkpoint in PyTorch is implemented by defining an autograd function which computes as usual function in the forward pass without storing intermediate activation maps but the inputs. In the backward pass, this function constructs a local computation graph for backward by recomputing the function using the stored inputs, and computes gradients by back-propagating through the local graph. However, this tightly binds $F_{i,j}'$ and $B_{i,j}$ together. Ultimately, we would like to insert the instruction for waiting the result $dx_i^{j}$ of $B_{i, j+1}$ to be copied from device $j+1$ to device $j$ in between $F_{i,j}'$ and $B_{i,j}$, to allow that $F_{i,j}'$ and the copy happens concurrently.

For such a fine-grained order control, {\torchgpipe} implements checkpointing with two separate autograd functions \cmtt{Checkpoint} and \cmtt{Recompute}. At the execution time of the task $F_{i, j}$, a pair of \cmtt{Checkpoint} and \cmtt{Recompute} which have a shared memory is generated. This shared memory is used in the backward pass for transferring the local computation graph made by executing \cmtt{Recompute} to \cmtt{ Checkpoint} for back-propagation. By arranging the functions so that $F_{i,j}'$, synchronization for receiving $dx_i^j$, and $B_{i,j}$ are executed in the order during the backward pass, it is ensured that re-computation and copy can happen concurrently.

\subsection{Dealing with Non-sequential Models}
\label{subsec:skip}

\begin{figure*}[b]
\center
    \includegraphics[width=\textwidth]{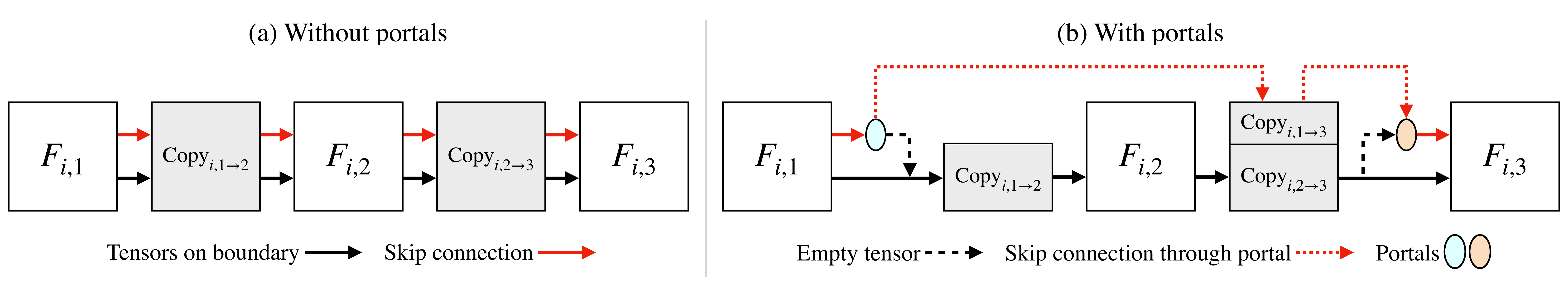}
    \caption{The flow of skip connection with or without portals. (a): Without portals, skipped tensor from device 1 is copied to device 2 and subsequently to device 3. (b): With portals, the tensor is directly copied to device 3. The gradient flows in the exact reverse direction in the backward pass.}
    \label{fig:portals}
\end{figure*}

In \autoref{sec:pipeline-parallelism}, we assumed that the model $f$ is composed of partitions $f^1, \cdots, f^n$ in sequence. In principle, any neural network can be represented in this form by sorting all nodes in the forward computation graph of $f$ in topological ordering. Hence, pipeline parallelism is applicable to any model.

However, consider a symptomatic case that all the partitions except the first and the last one are parallel, \emph{i.e.}, 
\[
f(y) = g^n(x^2, \cdots, x^{n-1})
\]
where $x^1 = g^1(x)$ and $x^j = g^j(x^1)$ for $j=2,\cdots,n-1$. In a sequential form, this is equivalent to $f = f^n \circ \cdots \circ f^1$ such that
\[
\begin{aligned}
    &f^n(x^1, x^2, \cdots, x^{n-1}) & &:= g^n(x^2, \cdots, x^{n-1}), \\
    &f^j(x^1, \cdots, x^{j-1}) & &:= (x^1, \cdots, x^{j-1}, f^j(x^1)) \\
\end{aligned}
\]
for $j=2,\cdots,n-1$, and $f^1 = g^1$. In this case, it is quite inefficient to use pipeline parallelism in its native form since at the boundary of device $j-1$ and $j$, the tuple $(x_i^1, \cdots, x_i^{j-1})$ must be copied instead of a single tensor $x_i^1$ which is the only required data to compute $j$th partition. 

{\torchgpipe} provides a submodule which allows users to indicate skipping tensors from which layer to which layer: \cmtt{torchgpipe.skip}. With the decorator \cmtt{@skippable}, user-defined layer can stash a tensor for later or pop a stashed one via \cmtt{yield} operator in Python without returning it. This in particular does not change the input and output signature of a layer. Hence, minimal effort is needed for adding skip connection to a preexisting sequential model. 

\subsubsection{Hiding Skip Tensors in the Graph: Portals}
\label{subsubsec:portals}

Adding skip connections into the dependency graph (\autoref{fig:pp-with-checkpointing}) is fairly straightforward. Indeed, no additional dependency would be introduced no matter how many skip connections are added, hence only the copy kernels for skip connections need extra care. In {\torchgpipe}, this is taken care by \emph{portals} consisting of three autograd functions \cmtt{PortalBlue}, \cmtt{PortalOrange}, and \cmtt{PortalCopy} sharing memory, like \cmtt{Checkpoint} and \cmtt{Recompute} in \autoref{subsubsec:checkpointing}. Each does the job of saving the skip tensor, loading the tensor, and moving the saved tensor to the skipped device, respectively (and vice versa in the backward pass). This mechanism is illustrated in \autoref{fig:portals}.

%% file: sections/4_experiments.tex
\section{Experiments}
\label{sec:experiments}

Every experiment was conducted with NVIDIA Tesla P40 GPUs with CUDA 10.1.243, each having 22 GiB of memory. For reproducibility, codes for all benchmarks provided in this section is made available in the repository\footnote{Further details available at \href{https://torchgpipe.readthedocs.io/en/stable/benchmarks.html}{this link}.}.

\subsection{Effects of Optimization Components}
\label{subsec:ablation-study}

We conducted an experiment to show that every component of {\torchgpipe} is necessary to achieve the maximal efficiency. Starting from the baseline which only has deterministic clock-cycle but no others, each component (backward dependency via \cmtt{Fork} and \cmtt{Join}, non-default streams for copy kernels, and portals for skip connections) is added incrementally. We report the throughput, GPU utilization, and memory usage under each setting to measure how each component contributed to the performance of {\torchgpipe}. We find that addition of each component gives a speed-up, and with all components {\torchgpipe} runs nearly twice as fast as the baseline. Results can be found in \autoref{table:components-ablation}.

We used U-Net for the experiment. Details of the architecture can be found in \autoref{subsubsec:unet-memory} and we set $(B, C)$ to be $(5, 64)$ as in the speed benchmark. In settings without portals, the model is implemented as a fully sequential version where skip connections are encoded as inputs and outputs of layers that they pass through, as described in the symptomatic example of \autoref{subsec:skip}. For the setting with all components, it is implemented with \cmtt{torchgpipe.skip} while the architecture is identical. 

\begin{figure*}[t]
    \centering
    \includegraphics[width=\textwidth]{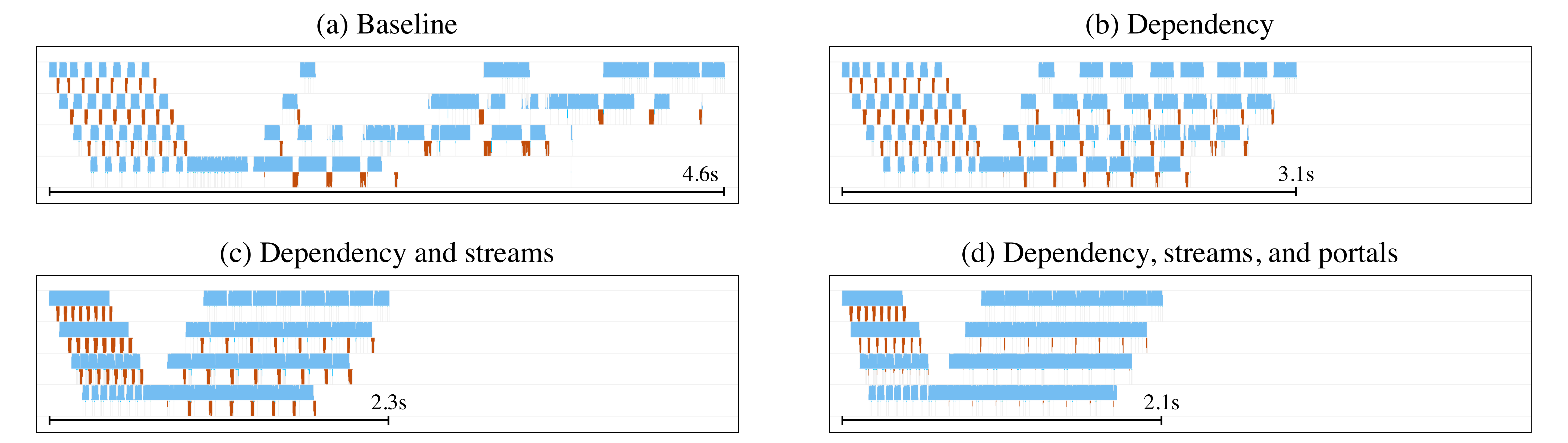}
    \caption{Detailed view of CUDA timeline for each setting in \autoref{table:components-ablation}, profiled with NVIDIA Nsight Systems 2019.5.1.58. Starting from the top, adjacent lanes with blue bars and red bars visualize the timeline per device. Blue bars represent computation kernels while red bars represent device-to-device copy (length proportional to time).}
    \label{fig:unet-timelines}
\end{figure*}

We also visualized per GPU timelines to help understanding each component's role, illustrated in \autoref{fig:unet-timelines}. Explanation for each picture is summarized as follows.

\begin{table*}[t]
    \centering
    \begin{tabular}{ccc|c|c|c|c}
        \multicolumn{3}{c|}{\bf{Optimization components}} & \bf{Throughput} & \bf{Speed up} & \bf{Utilization} & \bf{Memory usage} \\
    \hline
        $\times$   & $\times$ & $\times$ & 30.662/s &     1 & 44\% & 52.2 GiB \\
        Dependency & $\times$ & $\times$ & 41.306/s & 1.347 & 59\% & 19.1 GiB \\
        Dependency & Streams  & $\times$ & 55.191/s & 1.800 & 71\% & 30.0 GiB \\
        Dependency & Streams  & Portals  & 58.477/s & 1.907 & 75\% & 23.5 GiB \\
    \end{tabular}
    \caption{Performance of {\torchgpipe} when optimization components are incrementally added. The U-Net model with $(B, C)=(5, 64)$ is used for the experiment. The batch size and the number of micro-batches are fixed as 128 and 8, respectively. The model is partitioned and placed on four devices via {\torchgpipe}. Here the partition was found manually with the aid of \cmtt{torchgpipe.balance}.}
    \label{table:components-ablation}
\end{table*}

\begin{itemize}
    \item[(a)] By deterministic clock-cycle, all kernels are issued in the correct order during forward pass. It is illustrated by the left part of the timeline. However, without explicit dependency encoded in the computation graph, the autograd engine processes the micro-batches in an uncontrollable order so the timeline is messed up.
    \item[(b)] With backward dependency, kernels are now issued in the correct, deterministic order in backward pass.
    \item[(c)] By using non-default copy streams, copies and computations are now concurrent as illustrated by overlapping blue and red bars.
    \item[(d)] Portals remove unnecessary copies caused by transferring the skipping tensor to all devices in between. This is illustrated by that the length of red bars are reduced compared to (c). 
\end{itemize}

\subsection{Performance Benchmarks}

To demonstrate the efficiency of {\torchgpipe}, we report performance benchmarks similar to that conducted by GPipe \cite{huang2019gpipe}. 




\subsubsection{AmoebaNet-D Speed Benchmark}
\label{subsubsec:amoebanetd-speed}
We measured the throughput of AmoebaNet-D with various number of devices. For this, we measured the throughput of the model when {\torchgpipe} is applied, with $n$ partitions and $m$ micro-batches. Here throughput means the number of samples processed per second. 

The experiment is conducted for each pair $(m, n)$ where $m \in \{1, 4, 32\}$ and $n \in \{2, 4, 8\}$. When $m=1$, we used checkpointing to all micro-batches\footnote{{\torchgpipe} does not use checkpointing on the last micro-batch by default, as explained in \autoref{sec:pipeline-parallelism}. This means that no checkpointing is applied when $m=1$.} to make a fair comparison of loss due to checkpointing with \cite{huang2019gpipe}. The model we used is our implementation of a sequential version of AmoebaNet-D in PyTorch\footnote{We tried to make it as close as possible to the model in the official repository of TensorFlow (\href{https://github.com/tensorflow/tpu/tree/master/models/official/amoeba_net}{link}).}.

The model is trained by plain SGD for 10 epochs and reported the average throughput over the epochs except the first one. To exclude the overhead caused by data loading, we used a synthesized dataset which consists of 10,000 images whose dimension is $3 \times 224 \times 224$. For each setting, the batch size and the number of micro-batches are chosen to maximize the throughput. Relative speed-up is calculated against the baseline case $(m, n) = (1, 2)$ and reported in \autoref{table:amoebanetd-speed}. We included the speed-up of GPipe for comparison. 

The relative speed-up of {\torchgpipe} shows similar trend to that of GPipe. We remark that differences in performance reported in \autoref{table:amoebanetd-speed} might be due to many unknown factors such as balance of the partitions, discrepancy between the implementation, difference in devices, and so on.

\subsubsection{U-Net Memory Benchmark}
\label{subsubsec:unet-memory}

To evaluate the effectiveness of {\torchgpipe} for models with long skip connections, we used U-Net \cite{RonnebergerFB15} for 2-dimensional segmentation. The version of U-Net we used has five down-sampling layers and five up-sampling layers, and two hyper-parameters $B$ and $C$ determining the size of the model. Here $B$ stands for the number of convolution blocks in between down-sampling layers, and $C$ stands for the number of output channels of the first convolution. Channels are doubled after each down-sampling layers (or halved after each up-sampling layers, respectively). Our implementation of U-Net is rather symmetric than the original model proposed in \cite{RonnebergerFB15} for effective balancing.

We conducted an experiment to measure the ability of {\torchgpipe} for training a bigger model. For 1, 2, 4 and 8 GPUs, we found maximum $(B, C)$ to occupy each number of devices. In all settings, the input size is set to $3 \times 192 \times 192$, the output size to $1 \times 192 \times 192$, and the batch size to 32. The total memory usage for training each model is reported in \autoref{table:unet-memory}. Here parameters consumes 8 bytes each for itself and its gradients.

\subsubsection{U-Net Speed Benchmark}
\label{subsubsec:unet-speed}
We also measured the throughput of U-Net with various number of devices. Naive-1 denotes the baseline without pipeline parallelism nor checkpointing, and Pipeline-1, -2, -4, -8 denotes that the model is trained with {\torchgpipe} with the corresponding number of partitions. The hyper-parameters determining the size of U-Net is set to $(B, C) = (5, 64)$ in this experiment. The batch size, the number of micro-batches ($m$), and the balance to partitions are chosen to maximize the throughput. For each setting, throughput is measured as in \autoref{subsubsec:amoebanetd-speed} except that the image size was $3 \times 192 \times 192$ in this experiment. Result is summarized in \autoref{table:unet-speed}.

\begin{table}[t!]
    \small
    \centering
    \begin{tabular}[t]{c|ccc|ccc}
        \bf{AmoebaNet-D} & \multicolumn{3}{c|}{GPipe \cite{huang2019gpipe}} & \multicolumn{3}{|c}{Ours} \\
    \hline
        $n =$ & 2 & 4 & 8 & 2 & 4 & 8 \\
    \hline
        $m = 1$  &    1 & 1.13 & 1.38 &    1 & 1.00 & 0.93 \\
        $m = 4$  & 1.07 & 1.26 & 1.72 & 1.54 & 1.67 & 2.62 \\
        $m = 32$ & 1.21 & 1.84 & 3.48 & 1.77 & 2.71 & 4.95 \\
    \end{tabular}
    \caption{Speed benchmark on AmoebaNet-D (18, 256). In \cite{huang2019gpipe}, Cloud TPUv3s were used while we used NVIDIA Tesla P40 GPUs in our experiments.}
    \label{table:amoebanetd-speed}
    \vspace{10pt}
    \begin{tabular}[t]{c|c|c|c}
        \bf{U-Net} & ($B$, $C$) & Parameters & Memory usage \\
    \hline
        Naive-1    & (6, 72)   & 362.2M &  20.3 GiB \\
        Pipeline-1 & (11, 128) &  2.21B &  20.5 GiB \\
        Pipeline-2 & (24, 128) &  4.99B &  43.4 GiB \\
        Pipeline-4 & (24, 160) &  7.80B &  79.1 GiB \\
        Pipeline-8 & (48, 160) & 15.82B & 154.1 GiB
    \end{tabular}
    \caption{Memory benchmark on U-Net.}
    \label{table:unet-memory}
    \vspace{10pt}
    \begin{tabular}[t]{c|c|c|c|c}
        \bf{U-Net} & Throughput & Speed up & Batch size & $m$ \\
    \hline
        Naive-1    & 28.500/s &     1 &  40 & $\times$ \\
        Pipeline-1 & 24.456/s & 0.858 &  80 &        2 \\
        Pipeline-2 & 35.502/s & 1.246 & 512 &       32 \\
        Pipeline-4 & 67.042/s & 2.352 & 512 &       16 \\
        Pipeline-8 & 88.497/s & 3.105 & 640 &       40
    \end{tabular}
    \caption{Speed benchmark on U-Net with $(B, C)=(5, 64)$.}
    \label{table:unet-speed}
\end{table}

%% file: sections/5_conclusion.tex
\section{Conclusion}
\label{sec:conclusion}

In this paper, we introduced {\torchgpipe}, a ready-to-use library in PyTorch for micro-batch pipeline parallelism with checkpointing proposed by GPipe \cite{huang2019gpipe}. This library is designed and implemented in PyTorch's define-by-run and eager execution environment. Ablation study and performance benchmarks presented in \autoref{sec:experiments} demonstrate that all components of {\torchgpipe} are essential to endeavor the desired advantanges of pipeline parallelism with checkpointing in eager execution environment. We believe that general principles we established in the paper apply to any other frameworks with eager execution environment.

We tried to avoid going too deep into technical details involved in {\torchgpipe}. Our code is available at \url{https://github.com/kakaobrain/torchgpipe} for those who are interested in further details, and those who want to apply pipeline parallelism to their model in PyTorch.





